\documentclass[a4paper,11pt]{article}

\usepackage[utf8]{inputenc}
\usepackage[T1]{fontenc}
\usepackage{lmodern}
\usepackage{geometry}
\usepackage{graphicx}
\usepackage{authblk}
\usepackage{url}
\usepackage{cite}
\usepackage{fancyhdr}
\usepackage[hidelinks]{hyperref}
\usepackage{orcidlink}
\usepackage{subcaption}

\fancypagestyle{firstpage}{
	\fancyhf{}
	\fancyhead[C]{\includegraphics[height=2.2cm]{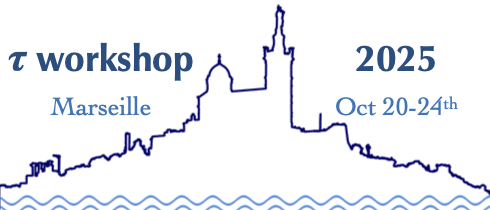}\\ \textit{Proceedings of the 18th International Workshop on Tau Lepton Physics}
	} 

}

\title{Generalized structure functions in semileptonic tau decays}
\author[1]{Daniel A. López Aguilar\,\orcidlink{0009-0002-5872-0062}}

\author[2]{Antonio Rodríguez Sánchez\,\orcidlink{0000-0001-7291-2146}}

\author[1]{Pablo Roig\,\orcidlink{0000-0002-6612-7157}\thanks{Presenting author.}}

\author[3]{Hanchen Yu\,
}

\affil[1]{Departamento de Física, Centro de Investigación y de Estudios Avanzados del Instituto
Politécnico Nacional Apartado Postal 14-740, 07360 Ciudad de México, México}
\affil[2]{Departamento de Física, Universidad de Castilla-La Mancha, Avenida de Carlos III,
s/n, 45004 Toledo, Spain}
\affil[3]{Departament de Física Teòrica, Instituto de Física Corpuscular, Universitat de
València — Consejo Superior de Investigaciones Científicas, Parc Científic, Catedrático
José Beltrán 2, E-46980 Paterna, Valencia, Spain}

\date{}

\begin{document}

\maketitle
\thispagestyle{firstpage}

\begin{abstract}
In case there are tensor interactions beyond the SM in the low-energy effective Lagrangian, their interference with the SM V-A currents yields a contribution for tau decays into three or more mesons that cannot be included in the famous structure functions introduced by Kühn and Mirkes in their seminal 1992 paper. We present this generalization here, highlighting again the importance of measuring the (generalized) spectral functions, as being model-independent and exploiting maximally the data. This can be particularly relevant for CP and T violation studies with at least three mesons in the final state of the tau decay.
\end{abstract}

\section{Introduction}
Tau decays into hadronic final states have long been considered as privileged scenarios to learn about low- and intermediate-energy QCD (including resonance properties) in the rather clean conditions provided by the lepton decay. In addition to this, if strong interaction effects and radiative corrections are well under control, they also constitute an ideal arena to test the Standard Model (SM) and look for new physics effects~\cite{Pich:2013lsa,Rodriguez-Sanchez:2025nsm}.

K\"uhn and Mirkes defined and characterized the Structure Functions (SFs) in ref.~\cite{Kuhn:1992nz} (see also \cite{Kuhn:1991cc, Kuhn:1996dv}), giving the most general description of the hadron system in semileptonic tau decays within the SM (results also apply to spin-zero charged weak currents and to (axial)-vector currents with different strength and/or relative weight than in the SM).

However, antisymmetric tensor currents appear in the basis for effective $\tau^-\to\nu_\tau \bar{u} D \,(D=d,s)$ transitions~\cite{Cirigliano:2009wk} and can be incorporated into Chiral Perturbation Theory~\cite{Cata:2007ns}, so they also need to be covered by the SF formalism. Ref.~\cite{Arteaga:2022xxy} noted that, in three-meson decays, the interference between this tensor current and the (axial)-vector current(s) breaks the factorization between the lepton and hadron tensors in the original SFs, calling for a generalization, which is the focus of this work~\cite{DanielThesis,WiP}.

Phenomenologically, the interest in antisymmetric tensor currents grew considerably as a possible explanation~\cite{Devi:2013gya} of the BaBar anomaly in the CP violation asymmetry in $\tau\to\nu K_S\pi^\pm$ decays~\cite{BaBar:2011pij}. However, effective field theory methods showed that only extreme fine-tuning could explain this signal given the constraints from neutral $D$ meson mixing and the neutron electric dipole moment~\cite{Cirigliano:2017tqn}. This accounts for uncertainties related to inelastic effects~\cite{Rendon:2019awg} and considers also~\cite{Chen:2019vbr} the null binned Belle measurement of this asymmetry~\cite{Belle:2011sna}, compatible with the BaBar anomaly. This puzzle is not settled yet, and modes with an additional $\pi^0$ or with a $K^\pm$ instead of the $\pi^\pm$ could help check the consistency of the experimental results~\cite{Aguilar:2024ybr,Grossman:2025uwz}.

Besides this BaBar anomaly, tensor currents give rise to novel and distinct sources of CP and T violation in meson tau decays, demanding thus a general formalism to guide the corresponding experimental searches for the decay channels with three (or more) mesons. The possibility to perform these analyses, including for the first time all possible low-energy directions allowed by the EFT description~\cite{Cirigliano:2009wk}, at Belle-II~\cite{Belle-II:2018jsg} and future facilities, like the super-charm tau factory~\cite{Cheng:2025kpp}, motivates our research.

\section{Main results}
The SF formalism~\cite{Kuhn:1992nz} introduces two convenient coordinate systems, \(S\) and \(S'\), both defined in the hadronic rest frame. The choice of axes in \(S\) simplifies the hadronic tensor, while \(S'\) is convenient for describing the \(\tau\) momentum and polarization. The rotation from \(S'\) to \(S\) is parameterized by the Euler angles \(\alpha,\,\beta,\,\gamma\).
While both $\beta$ and $\gamma$ are observable even if the $\tau$ direction cannot be reconstructed as a consequence of the undetected neutrino (like in B-factories), $\alpha$ is only measurable in super-tau-charm or similar environments. Other two angles, $\theta$ and $\psi$, characterize the relative orientation of the lab, hadron and rest frames, and are known in terms of measurable parameters~\cite{Kuhn:1992nz}. For unpolarized electron beams, $\tau$ polarization in the lab frame is small in Z decays and negligible at meson factories. We will sum and average over $\tau$ polarizations in the following.

Barring antisymmetric tensor currents, the most general description of the squared matrix element ($|\mathcal{M}|^2$) can be decomposed in the sum of 16 products of lepton and hadron functions ($|\mathcal{M}|^2=L_{\mu\nu}H^{\mu\nu}=\sum_{X}L_XW_X$), the latter being the SFs, which depend only on form factors and on hadron variables, unlike the lepton tensor. Although $\beta$ and $\gamma$ are measurable, it is convenient to use appropriate weights to integrate $|\mathcal{M}|^2$ over them and disentangle some SFs~\footnote{For instance, the spectral functions with spin $0$ and $1$ are proportional to the SFs $W_{SA}$ and $W_A+W_B$, respectively, with the full decay width proportional to $W_{SA}+\left(1+\frac{2Q^2}{M_\tau^2}\right)(W_A+W_B)$~\cite{Kuhn:1992nz}.}, a property that we will also exploit. We will benefit as well from the fact that (very approximate) SM symmetries, like isospin or G-parity, simplify the analysis for given decay channels.

In particular, these symmetries forbid an antisymmetric tensor current coupling to the $(3\pi)^-$ system, so we will focus on the $\tau^-(P)\to\eta^{(\prime)}(p_1)\pi^-(p_2)\pi^0(p_3)\nu_\tau(p)$ decays, where the hadron system couples to the tensor current. In the SM, using again these symmetries, only the vector current contributes, giving rise to a single form factor~\cite{GomezDumm:2012dpx}
\begin{equation}\label{HmuSM}
H_{\rm SM}^\mu=\langle\eta^{(\prime)}\pi^-\pi^0|\bar{d}\gamma^\mu u|0\rangle=iF_3^V(Q^2,s_1,s_2)\epsilon^{\mu\alpha\beta\gamma}p_{1\alpha}p_{2\beta}p_{3\gamma}\,,
\end{equation}
with $Q=p_1+p_2+p_3,\,s_i=(Q-p_i)^2$. As $F_3$, the $F_{1,2,4}$ form factors, stemming from the axial current, are also defined following ref.~\cite{Kuhn:1992nz}.

We recall that the most general description of semi-leptonic tau decays is given by the effective Lagrangian~\cite{Cirigliano:2009wk} (we take $D=d$, according to the preceding discussion, and omit this index below)
\begin{eqnarray}
\mathcal{L}&=&-\frac{G_F\widetilde{V_{ud}}}{\sqrt{2}}\Bigg\lbrace \bar{\tau}\gamma^\mu(1-\gamma_5)\nu_\tau\cdot\Big[\bar{u}\gamma_\mu(1-\gamma_5)d+\bar{u}\gamma_\mu(\epsilon^\tau_V+\epsilon^\tau_A\gamma_5)d\Big]\nonumber\\
&+&\bar{\tau}(1-\gamma_5)\nu_\tau\cdot \bar{u}(\epsilon^\tau_S-\epsilon^\tau_P\gamma_5)d+2\epsilon_T^\tau \bar{\tau}\sigma^{\mu\nu}(1-\gamma_5)\nu_\tau\cdot \bar{u}\sigma_{\mu\nu}d\Bigg\rbrace + \mathrm{h.c.},
\end{eqnarray}
with $G_F V_{ud}=G_F \widetilde{V_{ud}}(1+\epsilon_V^e)$. We do not display the $\tau$ upper index on the $\epsilon_{T}$ Wilson coefficients in what follows.

At leading order in the chiral counting, tensor interactions induce the hadron matrix element~\cite{Arteaga:2022xxy}
\begin{equation}
H_T^{\mu\nu}=i\frac{\Lambda_2C_q^{(\prime)}}{\sqrt{2}F^3}\epsilon^{\mu\nu\alpha\beta}(p_{3\alpha}p_{2\beta}-p_{2\alpha}p_{3\beta})\equiv \widehat{F}_T(0,0,0) \epsilon^{\mu\nu\alpha\beta}(p_{3\alpha}p_{2\beta}-p_{2\alpha}p_{3\beta})\,,
\end{equation}
where the normalization is determined from lattice QCD and $\eta-\eta^\prime$ mixing~\cite{Baum:2011rm,ParticleDataGroup:2024cfk}. The energy-dependence of the tensor form factor, $ \widehat{F}_T(Q^2,s_1,s_2)$, is built following refs.~\cite{GomezDumm:2012dpx,Miranda:2018cpf,Arteaga:2022xxy}.

Using self-duality is advantageous for dealing with the contraction among the $H^{\mu\nu}_TH^{\eta\dagger}_{V,A}$ product and its corresponding lepton counterpart. This explains why there are only $12$ generalized SFs, instead of $64$. In particular, we find~\cite{DanielThesis,WiP}($-\mathrm{Tr}(L^I H^I)$ is also used to group the last two terms below, which encode $6$ SFs~\footnote{${H}_S$ stands for the symmetrized traceless part of the tensor $H^I$ and the quantity $\tau_H$ for its trace, and analogously for the lepton part.})
\begin{equation}
|\mathcal{M} |^2= \dots + \sum_{\rm pols} 2 \epsilon_{T} L_{\mu \nu}H^{\mu \nu} L_{\eta}^{\dagger}H^{\dagger \eta} + \mathrm{h.c.} = \dots + \vec{L}_{0}\cdot\vec{H}_{0}-\vec{L}_A\cdot \vec{H}_{A} -\tau_L \, \tau_H - \mathrm{Tr}(L_{S}H_{S}) \, ,
\end{equation}
where, in general, (see ref.~\cite{DanielThesis,WiP} for the lepton part and other useful results
)
\begin{eqnarray}\label{generalizedSFS}
    \vec{H}_0 &&= \sum_{{i,j=1,2,3  \above 0pt i<j }}i\left(\sqrt{Q^2}F_4^*\widehat{F}_T\left[E_i\vec{p_j}-E_j \vec{p_i} + i\left(\vec{p_i}\times \vec{p_j}\right)\right]\right), \\
    \vec{H}_{A} &&= \sum_{{i,j=1,2,3  \above 0pt i<j }}i\widehat{F}_T\left(E_{i}\left(\vec{p}_{j}\times\vec{H}^{\dagger}\right)- E_{j}\left(\vec{p}_{i}\times\vec{H}^{\dagger}\right) + \mathit{i}\left[\vec{P}_{j}\left(\vec{p}_{i}\cdot\vec{H}^{\dagger}\right)- \vec{P}_{i}\left(\vec{p}_{j}\cdot\vec{H}^{\dagger}\right)\right]\right) ,\nonumber \\
    H^{I} &&= \sum_{{i,j=1,2,3  \above 0pt i<j }}i\,
    \widehat{F}_T \left(E_{i}\left[\vec{p}_{j}\otimes\vec{H}^{\dagger}\right] _S- E_{j}\left[\vec{p}_{i}\otimes\vec{H}^{\dagger}\right]_S + \mathit{i}\left[(\vec{p}_{i}\times\vec{p}_{j})\otimes\vec{H}^{\dagger}\right]_S\right)\,. \nonumber  
\end{eqnarray}
In eq.~(\ref{generalizedSFS}), $E_i$ are the meson energies in the hadron rest frame and $\vec{H}$ contains the spacelike components of $H_{\rm SM}^{\dagger\eta}$, eq.~(\ref{HmuSM}).

As anticipated, use of moments proves convenient to isolate SFs. A moment $m$ is defined as
\begin{equation}\label{defmoment}
\langle m\rangle = \int  \sin\beta d\beta\, d\alpha \, d\gamma \,m(\alpha,\beta,\gamma)\, \frac{1}{\Gamma_\tau}  \frac{d^7\Gamma}{dQ^2ds_1ds_2d\theta d\alpha d\beta d\gamma}\,.
\end{equation}
We illustrate this procedure only with a few results~\cite{DanielThesis,WiP}, involving the $z$-component (${}^3$) of $\vec{H}_{0,A}$,
\begin{eqnarray}
\left\langle \sin \beta\sin\gamma \right\rangle &=& \frac{1}{2}\left((\xi_+-\xi_-)H_0^3+(\zeta_+-\zeta_-)H_A^3 + \dots
    \right)    +\rm{c.c.},\nonumber \\
\left\langle \sin \beta\cos\gamma \right\rangle &=& \frac{1}{2}\left((\xi_++\xi_-)H_0^3+(\zeta_++\zeta_-)H_A^3+\dots
    \right)    +\mathrm{c.c.} ,\nonumber \\ 
     \left\langle \cos \beta\right\rangle &=& \epsilon_3\,H_0^3 + \zeta_3\,H_A^3 +\dots 
    +\mathrm{c.c.}, 
\end{eqnarray}
with known coefficients~\cite{DanielThesis,WiP}.

\begin{figure}[h]
    \centering
    \begin{subfigure}{0.49\textwidth}
        \includegraphics[width=0.9\linewidth]{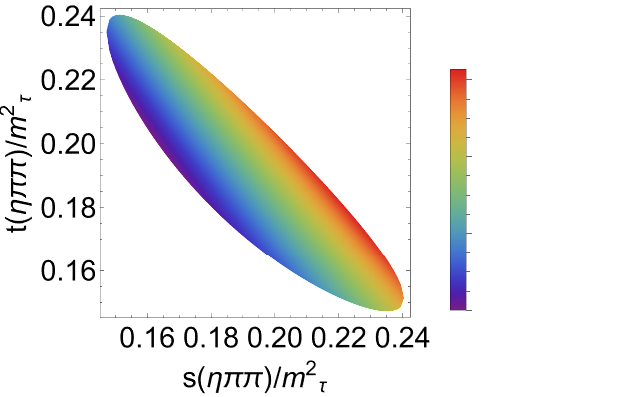}
        \label{fig_ia}
    \end{subfigure}
    \hfill
    \begin{subfigure}{0.49\textwidth}
        \includegraphics[width=0.9\linewidth]{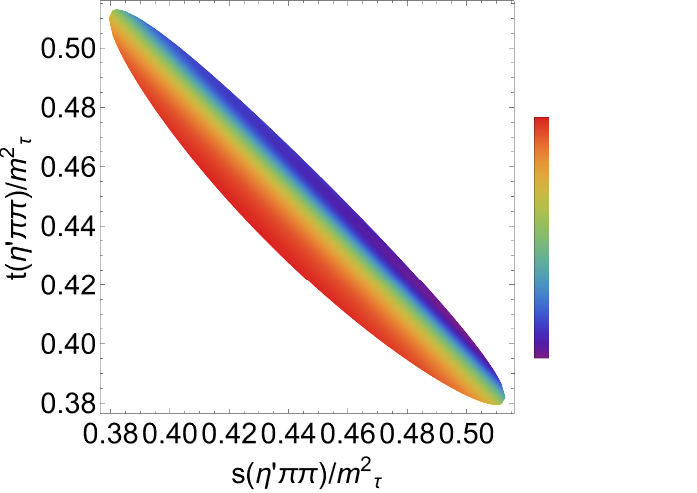}
        \label{fig_ib}
    \end{subfigure}
    \caption{Dalitz plot for the CP asymmetry distribution, eq.~(\ref{eq_ACP}), with $ \Re e\{\epsilon_T\} = 5\cdot 10^{-3}\,   $ and $ \Im m\{\epsilon_T \} = 8\cdot\,10^{-5} $~\cite{Cirigliano:2021yto} and a fixed $Q^2= (1000)^2$ MeV$^2$, for the $\tau \rightarrow \eta\pi\pi\nu_{\tau} $ channel (left). Same for the $\eta^\prime$ channel, with $Q^2=(1400)^2$ MeV$^2$ (right).}
    \label{fig_i}
\end{figure}

Using these generalized SFs, $\vec{H}_0,\vec{H}_A,\tau_H,H_S$, we have considered observables violating either CP or T symmetries, as well as null (CP and T conserving) observables that can be generated only if $\epsilon_T\neq0$. We exemplify this with the same observable shown in the TAU25 talk, which corresponds to a CP violating asymmetry, defined as
\begin{eqnarray}\label{eq_ACP}
    A_{CP}(Q^2,s1,s2) =2 \frac{ \frac{d^3\Gamma}{dQ^2\,ds1\,ds2}(\tau^{-} \rightarrow \eta^{(\prime)}\pi^-\pi^0\nu_{\tau}) - \frac{d^3\Gamma}{dQ^2\,ds1\,ds2}(\tau^{+} \rightarrow \eta^{(\prime)}\pi^+\pi^0\bar{\nu}_{\tau})  }  { \frac{d^3\Gamma}{dQ^2\,ds1\,ds2}(\tau^{-} \rightarrow \eta^{(\prime)}\pi^-\pi^0\nu_{\tau}) + \frac{d^3\Gamma}{dQ^2\,ds1\,ds2}(\tau^{+} \rightarrow \eta^{(\prime)}\pi^+\pi^0\bar{\nu}_{\tau}) }  ,
\end{eqnarray}
where the numerator is clearly proportional to $\Im m \epsilon_T\Im m\left(\widehat{F}_TF_3^V\right)$, in excellent approximation, and the denominator is dominated by the SM contribution. Our preliminary results for this asymmetry (for fixed $Q^2$ values) are shown in fig.~\ref{fig_i}. Relative differences between the largest and smallest $A_{CP}$ values within the Dalitz plot reach $\sim 40\%$ ($\eta$ case) and $\sim200\%$ ($\eta^\prime$ channel).

\section{Conclusions}
We first want to stress the usefulness of the Kuhn-Mirkes SFs to analyze semileptonic tau decays, including tau polarization. They allow for a model-independent description of hadron information and their measurement should therefore be a priority in current and forthcoming experiments. Lacking enough data to completely characterize them, integration over some variables can still yield very valuable information.

Here we have focused on the interference between an antisymmetric tensor (appearing in the basis for the low-energy effective interactions) and a combination of vector and axial currents in three-meson tau decays, which is not covered by the original formalism. We have proposed a convenient generalization, encompassing this case (of particular importance in CP and T violation analyses) and consider its application to B- and super-tau-charm factories~\cite{DanielThesis,WiP}.

Altogether, the (generalized) structure functions continue to be precious tools, both in characterizing resonance dynamics and in new physics searches, as we have illustrated. Although they were published by ALEPH~\cite{ALEPH:1998rgl}, CLEO~\cite{CLEO:1999maq} and OPAL~\cite{OPAL:1995txk,OPAL:1997was}~\footnote{These data constrained theory at the same level than the spectrum, however~\cite{Dumm:2009va,Roig:2010csr}.}, later experiments have not measured them. We hope that our work motivates our experimental colleagues to undertake their search.

\section*{Acknowledgements}
DALA thanks Secihti funding his Ph. D. ARS was supported by the Generalitat Valenciana
(Spain) through the plan GenT program (CIDEIG/2023/12). PR acknowledges the support given by Secihti through the project CBF2023-2024-3226 and is indebted to the organizers for support and this great workshop! HY receives support from the Generalitat Valenciana (Spain) through the plan GenT program CIDEGENT/2021/037. This work has also been supported by 
the Spanish Government (Agencia Estatal de Investigaci\'on MCIN/AEI/10.13039/501100011033) Grant No. PID2023-146220NB-I00.
\bibliographystyle{JHEP}
\bibliography{references}

\end{document}